\documentstyle[12pt]{article}
\begin{document}
\setcounter{footnote}{0}
\renewcommand{\thefootnote}{\alph{footnote}}
\renewcommand{\theequation}{\thesection.\arabic{equation}}
\newcounter{saveeqn}
\newcommand{\add}{\addtocounter{equation}{1}}
\newcommand{\alpheqn}{\setcounter{saveeqn}{\value{equation}}%
\setcounter{equation}{0}%
\renewcommand{\theequation}{\mbox{\thesection.\arabic{saveeqn}{\alph{equation}}}}}
\newcommand{\reseteqn}{\setcounter{equation}{\value{saveeqn}}%
\renewcommand{\theequation}{\thesection.\arabic{equation}}}
\newenvironment{nedalph}{\add\alpheqn\begin{eqnarray}}{\end{eqnarray}\reseteqn}
\newsavebox{\PSLASH}
\sbox{\PSLASH}{$p$\hspace{-1.8mm}/}
\newcommand{\PS}{\usebox{\PSLASH}}
\newsavebox{\PARTIALSLASH}
\sbox{\PARTIALSLASH}{$\partial$\hspace{-2.3mm}/}
\newcommand{\PARTIALS}{\usebox{\PARTIALSLASH}}
\newsavebox{\ASLASH}
\sbox{\ASLASH}{$A$\hspace{-2.1mm}/}
\newcommand{\AS}{\usebox{\ASLASH}}
\newsavebox{\QSLASH}
\sbox{\QSLASH}{$q$\hspace{-2.1mm}/}
\newcommand{\QS}{\usebox{\QSLASH}}
\newsavebox{\KSLASH}
\sbox{\KSLASH}{$k$\hspace{-1.8mm}/}
\newcommand{\KS}{\usebox{\KSLASH}}
\newsavebox{\LSLASH}
\sbox{\LSLASH}{$\ell$\hspace{-1.8mm}/}
\newcommand{\LS}{\usebox{\LSLASH}}
\newsavebox{\LLSLASH}
\sbox{\LLSLASH}{$L$\hspace{-1.8mm}/}
\newcommand{\LLS}{\usebox{\LLSLASH}}
\newsavebox{\SSLASH}
\sbox{\SSLASH}{$s$\hspace{-1.8mm}/}
\newcommand{\SS}{\usebox{\SSLASH}}
\newsavebox{\DSLASH}
\sbox{\DSLASH}{$D$\hspace{-2.8mm}/}
\newcommand{\DS}{\usebox{\DSLASH}}
\newsavebox{\DbfSLASH}
\sbox{\DbfSLASH}{${\mathbf D}$\hspace{-2.8mm}/}
\newcommand{\DBFS}{\usebox{\DbfSLASH}}
\newsavebox{\DELVECRIGHT}
\sbox{\DELVECRIGHT}{$\stackrel{\rightarrow}{\partial}$}
\newcommand{\PARVECR}{\usebox{\DELVECRIGHT}}
\thispagestyle{empty}
\begin{flushright}
{\tt{hep-th/0302179}}
\end{flushright}
\vspace{1.5cm}
\begin{center}
{\large\bf{BRST Quantization of Noncommutative Gauge Theories}}\\
\vspace{1.5cm} {Masoud Soroush}\ \footnote{\normalsize{\
Electronic address:
soroush@mehr.sharif.edu}}  \\
\vspace{0.5cm}
{\sl  Department of Physics, Sharif University of Technology}\\
{\sl P.O. Box 11365-9161, Tehran-Iran}\\
\end{center}
\vspace{1cm}
\begin{center}
{\bf {Abstract}}
\end{center}
\vspace{0.1cm}
\begin{quote}
In this paper, the BRST symmetry transformation is presented for
the noncommutative $U(N)$ gauge theory. The nilpotency of the
charge associated to this symmetry is then proved. As a
consequence for the space-like non-commutativity parameter, the
Hilbert space of physical states is determined by the cohomology
space of the BRST operator as in the commutative case. Further,
the unitarity of the S-matrix elements projected onto the
subspace of physical states is deduced.
\end{quote}
\hspace{0.8cm}
\par\noindent
{\it PACS No.:} 11.15.Bt, 11.10.Gh, 11.25.Db
\par\noindent
{\it Keywords:} Noncommutative Gauge Theory, BRST Symmetry,
Nilpotent Charge.
\newpage
\setcounter{page}{1} \setcounter{equation}{0}
\section{Introduction}
In the past few years, a lot of work has been devoted to study of
noncommutative gauge theories. The main motivation for these
studies is related to the realization of such theories in the
framework of string theory. It turns out that noncommutative
supersymmetric gauge theories appear as the low energy effective
theory on a D-brane in the presence of a nonzero NS-NS two form
background field\cite{h22,h2,h3}. Apart from the string theory
realization, the analysis of quantum mechanical features of
noncommutative gauge theories is important from field theoretical
point of view.
\par
As it is clear, the BRST symmetry which makes explicit the fact
that the quantization is independent of a choice of a particular
gauge is a method to quantization of gauge theories. This
procedure includes two stages. The first stage is the
introduction of ghosts, as in the standard Faddeev-Popov method.
The theory included ghosts is then quantized in the usual way by
ignoring the gauge symmetry. Since ghosts violate
spin-statistics, the space of physical states of the quantum
theory has pseudo-Hermitian rather than Hermitian inner product.
The second stage, however, restores positivity of the inner
product defined in the space of states and also brings back the
gauge symmetry. In this process, a continuous global symmetry
transformation (the BRST transformation) is defined on the algebra
of local operators. The nilpotency of the charge associated to
this global symmetry is then proved. Ultimately, the Hilbert space
structure on the quantum states is determined by the cohomology
space of the BRST operator (the charge associated to the BRST
symmetry). Consequently, the induced inner product on the
cohomology space will be positive.
\par
In this paper, the BRST quantization procedure is followed for the
noncommutative $U(N)$ gauge theory. In Sect. 2, after gauge fixing
and introducing ghost fields, the full action of the theory is
considered. The BRST symmetry transformation is then found in
Sect. 3 in a way that the full action of the theory remains
invariant under it. In Sect. 4, however, it is shown that the
charge associated to the BRST symmetry is nilpotent.
\par
Sect. 5 includes several parts. First in Sect. 5.1, we discuss
the BRST symmetry leads to a conserved charge which commutes with
the Hamiltonian of the theory only for the space-like
non-commutativity although it is found for a general
non-commutativity parameter. In the next part, Sect. 5.2, we
prove that the BRST symmetry is preserved at quantum level. More
precisely, we have to show that the path integral measure of the
partition function of the theory remains invariant under the BRST
transformation. Arguments presented in the last part of this
section, Sect. 5.3, are restricted to the case of the space-like
non-commutativity. We will show the space of physical states is
determined by the cohomology space of the BRST operator just as
the commutative case. Finally, the unitarity of the S-matrix
elements projected onto the subspace of physical states is
proved. Sect. 6 is devoted to a brief discussion.
\section{Full Action of the Noncommutative Gauge Theory}
\setcounter{equation}{0} In this section, we fix our notation.
Meanwhile, we introduce the action of the noncommutative $U(N)$
gauge theory\footnote{\ It should be noticed that constructing
the noncommutative gauge theory with other famous matrix Lie
groups $SU(N)$, $SO(N)$ or $Sp(N)$ is not possible, since Lie
algebras of these groups are not closed under the Moyal
bracket\cite{h9,h1}.}. The classical action of the noncommutative
pure Yang-Mills theory is given by:
\begin{eqnarray}\label{A1}
S_{g}[A]=-\frac{1}{4}\int\ \mbox{Tr}(F_{\mu\nu}\star F^{\mu\nu})\
,
\end{eqnarray}
where $\star$ denotes the Moyal star product and curvature
$F^{a}_{\mu\nu}$ is defined as follows:
\begin{eqnarray}\label{A2}
F^{a}_{\mu\nu}=\partial_{[\mu}A^{a}_{\nu]}
-ig[A^{b}_{\mu}t^{b},A^{c}_{\nu}t^{c}]_{\star}^{a}\ .
\end{eqnarray}
In order to separate the effect of non-commutativity, we introduce
some notations. Defining $h^{abc}t^{a}\equiv t^{b}t^{c}$, we can
express $F^{a}_{\mu\nu}$ as:
\begin{eqnarray}\label{A3}
F^{a}_{\mu\nu}=\partial_{[\mu}A^{a}_{\nu]}-igh^{abc}A^{b}_{\mu}\star
A^{c}_{\nu}+igh^{acb}A^{c}_{\nu}\star A^{b}_{\mu}\ .
\end{eqnarray}
The structure constants $f^{abc}$ and totally symmetric $d^{abc}$
factors of $U(N)$ are introduced by:
\begin{eqnarray}\label{A4}
[t^{a},t^{b}]=if^{abc}t^{c}\ ,\hspace{2cm}
\{t^{a},t^{b}\}=d^{abc}t^{c}\ .
\end{eqnarray}
Expressing $h^{abc}$ with respect to $f^{abc}$ and $d^{abc}$
factors\footnote{\ We suppose that the Killing form has been
defined, whether the indices are up or down is irrelevant in the
orthonormal basis.}:
\begin{eqnarray}\label{A5}
h^{abc}=\frac{i}{2}f^{abc}+\frac{1}{2}d^{abc}\ ,
\end{eqnarray}
we immediately conclude:
\begin{eqnarray}\label{A6}
F^{a}_{\mu\nu}=\partial_{[\mu}A^{a}_{\nu]}+\frac{1}{2}gf^{abc}\{A^{b}_{\mu},A^{c}_{\nu}\}_{\star}
-\frac{i}{2}gd^{abc}[A^{b}_{\mu},A^{c}_{\nu}]_{\star}\ .
\end{eqnarray}
Now it is clear if we put $\theta=0$, the last term will vanish
and the commutative expression will be obtained for
$F^{a}_{\mu\nu}$.
\par
The matter field part includes the connection, defined by:
\begin{eqnarray}\label{A7}
D=\partial-igA\star\ \ ,
\end{eqnarray}
where we considered the fundamental representation of the matter
field with respect to the star product. The matter part of the
action is then given by:
\begin{eqnarray}\label{A8}
S_{mat}[A,\psi,\bar{\psi}]=\int\ \bar{\psi}\star(i\DS-m)\psi\ .
\end{eqnarray}
The two parts $S_{g}$ and $S_{mat}$ of the action are separately
invariant under the following gauge transformations of matter and
gauge fields:
\begin{eqnarray}\label{A9}
&&\psi\rightarrow U\star\psi\ , \\
&&A\rightarrow U\star A\star U^{-1}+\frac{i}{g}U\star\partial
U^{-1}\ ,
\end{eqnarray}
where $U\in U_{\star}(N)$. However, the infinitesimal gauge
transformations of the matter and gauge fields take the forms:
\begin{eqnarray}\label{A10}
&&\delta\psi=ig\omega^{a}t^{a}\star\psi\ ,\\
&&\delta
A^{a}=(D^{\mathsf{adj}}\omega)^{a}-igh^{abd}[A^{d},\omega^{b}]_{\star}\
,
\end{eqnarray}
where $\omega^{a}$'s are infinitesimal local parameters of the
gauge transformation and $D^{\mathsf{adj}}$ is the connection
associated to the commutative $U(N)$ gauge group in the adjoint
representation and is introduced by:
\begin{eqnarray}\label{A11}
(D^{\mathsf{adj}})^{ac}=\delta^{ac}\partial+gf^{abc}A^{b}\star\ \
.
\end{eqnarray}
Obviously for the case $\theta=0$, the second term in the r.h.s.
of Eq. (2.12) vanishes and $\delta A^{a}$ becomes the expression
of the commutative case.
\par
In order to quantize the theory with the gauge symmetry, we have
to follow the standard Faddeev-Popov procedure. More precisely,
to compute the partition function of the theory, we should
integrate over the quotient space
${\mathcal{F}\big/{\mathcal{G}}}$, where ${\mathcal{F}}$ and
${\mathcal{G}}$ are the space of field configurations and the
group associated to the gauge symmetry respectively. Imposing the
covariant gauge condition on the decomposition of unity, we find
the gauge fixing and ghost parts of the action as:
\begin{eqnarray}\label{A12}
&&S_{gf}[A]=-\frac{1}{2}\xi^{-1}\int\ (\partial\cdot A^{a})^{2}\
,\\
&&S_{gh}[c,\bar{c},A]=-\int\Big\{\bar{c}^{a}\star(\partial\cdot
D^{\mathsf{adj}}c)^{a}-igh^{abd}\bar{c}^{a}\star\partial^{\mu}[A^{d}_{\mu},c^{b}]_{\star}\Big\}\
,
\end{eqnarray}
where $c$ and $\bar{c}$ denote the ghost and the antighost fields.
Ultimately, the full action of the noncommutative gauge theory is
found as follows:
\begin{eqnarray}\label{A13}
S[A,\psi,\bar{\psi},c,\bar{c}]=S_{g}[A]+S_{mat}[A,\psi,\bar{\psi}]+S_{gh}[A,c,\bar{c}]+S_{gf}[A]\
.
\end{eqnarray}
\section{BRST Symmetry}
\setcounter{equation}{0} As in the commutative case, any gauge
fixing procedure destroys gauge invariance of the noncommutative
gauge theory. Therefore, in order to obtain a sensible
quantization, we must make sure that in the final result the
gauge symmetry is restored. The BRST quantization method provides
a powerful tool to achieve this aim. The essential element for
applying this method is first introducing a global symmetry
transformation whose associated charge is nilpotent. In this
section, we present the BRST symmetry transformation for the
noncommutative $U(N)$ gauge theory.
\par
The BRST symmetry is a global transformation (the parameter of the
transformation is global.) which leaves the full action of the
theory invariant. Since the pure gauge and matter parts of the
action are invariant under gauge transformations Eq. (\ref{A10})
and Eq. (2.12), the most natural BRST transformations which can be
considered for the gauge and matter fields are the ordinary gauge
transformations. Therefore, we introduce the BRST transformations
for the gauge field $A^{a}$ and the matter field $\psi$ as:
\begin{eqnarray}\label{B1}
&&\delta A^{a}=(D^{\mathsf{adj}}c)^{a}\delta\lambda-igh^{acb}[A^{b},c^{c}]_{\star}\delta\lambda\ ,\\
&&\delta\psi=-ig(c^{a}t^{a}\star\psi)\delta\lambda\ ,
\end{eqnarray}
where $\delta\lambda$ (the parameter of the transformation) is a
global Grassmann variable. It should be noticed that
$\delta\lambda$ is not necessarily infinitesimal. But since
$(\delta\lambda)^{2}=0$, Eq. (\ref{B1}) and Eq. (3.2) introduce
infinitesimal gauge transformations. Hence, we have:
\begin{eqnarray}\label{B2}
\delta\Big(S_{g}[A]+S_{mat}[A,\psi,\bar{\psi}]\Big)=0\ .
\end{eqnarray}
Now we are going to find the transformations of the ghost and
antighost fields in a way that the action is invariant under this
transformation. Therefore, we conclude:
\begin{eqnarray}\label{B3}
\delta S=\delta\Big(S_{gh}+S_{gf}\Big)=0\ .
\end{eqnarray}
Using Eq. (\ref{B1}), the variation of the action is then given
by:
\begin{eqnarray}\label{B5}
\delta S&=&\int
\Biggr\{-\Big(\delta\bar{c}^{a}-\xi^{-1}(\partial\cdot A^{a})
\delta\lambda\Big)\star\Big(\partial\cdot(D^{\mathsf{adj}}c)^{a}-igh^{abd}\partial^{\mu}
[A^{d}_{\mu},c^{b}]_{\star}\Big)\nonumber\\
&&\hspace{0.7cm}-\bar{c}^{a}\star\partial\cdot\Big[\delta\Big((D^{\mathsf{adj}}c)^{a}-
igh^{abd}[A^{d},c^{b}]_{\star}\Big)\Big]\Biggr\}\ ,
\end{eqnarray}
which will vanish, if we require that:
\begin{eqnarray}\label{B6}
&\delta\bar{c}^{a}=\xi^{-1}(\partial\cdot A^{a})\delta\lambda\ ,\\
&\delta\Big((D^{\mathsf{adj}}c)^{a}-igh^{abd}[A^{d},c^{b}]_{\star}\Big)=0\
.
\end{eqnarray}
The first one introduces the BRST transformation of the antighost
field and the other one will then introduce the BRST
transformation of the ghost field. Considering the first term of
Eq. (3.7), we obtain:
\begin{eqnarray}\label{B8}
\delta(D^{\mathsf{adj}}c)^{a}&=&\partial(\delta
c^{a})+gf^{abc}\delta
A^{b}\star c^{c}+gf^{abc}A^{b}\star\delta c^{c}\nonumber\\
&=&\partial(\delta c^{a})-gf^{abc}\Big(\partial
c^{b}+gf^{bde}A^{d}\star c^{e}\Big)\star
c^{c}\delta\lambda\nonumber\\
&&+ig^{2}f^{abc}h^{bed}[A^{d},c^{e}]_{\star}\star
c^{c}\delta\lambda+gf^{abc}A^{b}\star\delta c^{c}\ ,
\end{eqnarray}
where in the last line, we used Eq. (\ref{B1}). For the second
term of Eq. (3.7), however, we have:
\begin{eqnarray}\label{B9}
\delta[A^{d},c^{b}]_{\star}=[\delta
A^{d},c^{b}]_{\star}+[A^{d},\delta c^{b}]_{\star}\ ,
\end{eqnarray}
again using Eq. (\ref{B1}), the first term in the r.h.s. of the
above equality itself reads:
\begin{eqnarray}\label{B10}
[\delta
A^{d},c^{b}]_{\star}=-\{(D^{\mathsf{adj}}c)^{d},c^{b}\}_{\star}\delta\lambda+igh^{dfe}\Big\{[A^{e},c^{f}]_{\star}
,c^{b}\Big\}_{\star}\delta\lambda\ .
\end{eqnarray}
Now substitute Eq. (\ref{B8}), Eq. (\ref{B9}), and Eq. (\ref{B10})
for Eq. (3.7). Then as the first step in Eq. (3.7), considering
all terms including space-time derivatives, we find:
\begin{eqnarray}\label{B11}
\mbox{terms including derivative}&=&\partial(\delta
c^{a})-gf^{abc}\partial c^{b}\star
c^{c}\delta\lambda+igh^{abd}\{\partial c^{d},c^{b}
\}_{\star}\delta\lambda\nonumber\\
&=&\partial\Big(\delta c^{a}-\frac{1}{2}gf^{abd}c^{b}\star
c^{d}\delta\lambda+\frac{i}{4}gd^{abd}\{c^{d},c^{b}\}_{\star}\delta\lambda\Big)\
.
\end{eqnarray}
To obtain the last expression, we used Eq. (\ref{A5}). Notice that
to derive the above expression, there is no necessity to exchange
the position of fields. We only have exchanged the position of the
gauge group indices of fields. The remained terms not including
space-time derivatives of Eq. (3.7) are:
\begin{eqnarray}\label{B12}
&&-g^{2}f^{abc}f^{bde}A^{d}\star c^{e}\star
c^{c}\delta\lambda+ig^{2}f^{abc}h^{bed}[A^{d},c^{e}]_{\star}\star
c^{c}\delta\lambda+gf^{abc}A^{b}\star\delta c^{c}\nonumber\\
&&+ig^{2}h^{abd}f^{def}\{A^{e}\star
c^{f},c^{b}\}_{\star}\delta\lambda+g^{2}h^{abd}h^{dfe}\Big\{[A^{e},c^{f}]_{\star},c^{b}\Big\}_{\star}\delta\lambda
-igh^{abd}[A^{d},\delta c^{b}]_{\star}\ .
\end{eqnarray}
Now the claim is that Eq. (3.7) can ultimately be rewritten in
the form of Eq. (\ref{B16}). To arrive to this equation, we
present the way that leads only to one of its term. One can find
more details in appendix A. For instance, consider the second
term of the first line of the above expression. Using Eq.
(\ref{A5}), we can write it as follows:
\begin{eqnarray}\label{B13}
ig^{2}f^{abc}h^{bed}[A^{d},c^{e}]_{\star}\star
c^{c}\delta\lambda&=&-\frac{1}{2}f^{abc}f^{bed}A^{d}\star
c^{e}\star
c^{c}\delta\lambda+\frac{1}{2}g^{2}f^{abc}f^{bed}c^{e}\star
A^{d}\star
c^{c}\delta\lambda\nonumber\\
&&+\frac{i}{2}f^{abc}d^{bed}[A^{d},c^{e}]_{\star}\star
c^{c}\delta\lambda\ .
\end{eqnarray}
Adding the first term of the r.h.s. of the above equality to the
first term of Eq. (\ref{B12}), it immediately results:
\begin{eqnarray}\label{B14}
-\frac{1}{2}g^{2}f^{abc}f^{bde}A^{d}\star c^{e}\star
c^{c}\delta\lambda=-\frac{1}{2}g^{2}(f^{abd}f^{bce}+f^{abe}f^{bdc})A^{d}\star
c^{e}\star c^{c}\delta\lambda\ ,
\end{eqnarray}
where we used the Jacobi identity in the r.h.s. of the above
expression. The first term in the r.h.s. of the above equality is
exactly the second term appeared in the second line of Eq.
(\ref{B16}). Therefore, Eq. (3.7) yields:
\begin{eqnarray}\label{B16}
&&\partial\Big(\delta c^{a}-\frac{1}{2}gf^{abd}c^{b}\star
c^{d}\delta\lambda+\frac{i}{4}gd^{abd}\{c^{b},c^{d}\}_{\star}\delta\lambda\Big)\nonumber\\
&&+gf^{abc}A^{b}\star \Big(\delta
c^{c}-\frac{1}{2}gf^{cde}c^{d}\star
c^{e}\delta\lambda+\frac{i}{4}gd^{cde}\{c^{d},c^{e}\}_{\star}\delta\lambda\Big)\nonumber\\
&&-igh^{abd}\Big[ A^{d},\delta c^{b}-\frac{1}{2}gf^{bef}c^{e}\star
c^{f}\delta\lambda+\frac{i}{4}gd^{bef}\{c^{e},c^{f}\}_{\star}\delta\lambda\Big]_{\star}=0\
.
\end{eqnarray}
In order to obtain vanishing result for the l.h.s. of the above
equality, it is sufficient to require that:
\begin{eqnarray}\label{B17}
\delta c^{a}=\frac{1}{2}gf^{abd}c^{b}\star
c^{d}\delta\lambda-\frac{i}{4}gd^{abd}\{c^{d},c^{b}\}_{\star}\delta\lambda\
,
\end{eqnarray}
which introduces the BRST transformation of the ghost field.
Therefore, we have proved that the full action of the
noncommutative $U(N)$ gauge theory remains invariant under the
following transformation:
\begin{eqnarray}\label{B18}
&&\delta
A^{a}=(D^{\mathsf{adj}}c)^{a}\delta\lambda-igh^{acb}[A^{b},c^{c}]_{\star}\delta\lambda\
,\nonumber\\
&&\delta\psi=-ig(c^{a}t^{a}\star\psi)\delta\lambda\ ,\\
&&\delta\bar{c}^{a}=\xi^{-1}(\partial\cdot A^{a})\delta\lambda\
,\nonumber\\
&&\delta c^{a}=\frac{1}{2}gf^{abd}c^{b}\star
c^{d}\delta\lambda-\frac{i}{4}gd^{abd}\{c^{b},c^{d}\}_{\star}\delta\lambda\
.\nonumber
\end{eqnarray}
To compare the above result with the commutative case, the above
symmetry transformation takes its commutative form if we put
$\theta=0$. Further considering the $U(1)$ gauge group, we can
study the situation of the noncommutative QED. It is easy to see
that the above transformation leads to the BRST transformation
which was first found in \cite{h8}.
\section{Nilpotent Charge of the BRST Symmetry}
\setcounter{equation}{0} In this section, we will show that the
charge of the BRST symmetry is nilpotent. The charge $Q$
associated to the BRST symmetry satisfies the following relation:
\begin{eqnarray}\label{C}
\delta\phi=i[\delta\lambda\ Q,\phi\ ]\ ,
\end{eqnarray}
where $\delta\phi$ is introduced by the transformations Eq.
(\ref{B18}). Notice that the commutator appeared in Eq. (\ref{C})
is not the Moyal bracket. For simplicity, we define the action of
charge $Q$ on each field $\phi$ as:
\begin{eqnarray}\label{C1}
\delta\phi=\delta\lambda ({\mathcal{Q}}\phi)=i[\delta\lambda\
Q,\phi\ ]\ ,
\end{eqnarray}
where ${\mathcal{Q}}=i[Q,\ ]_{\pm}$ satisfies the super Leibnitz
rule. Considering the transformations Eq. (\ref{B18}), we find the
action of charge $Q$ on each field of the theory as follows:
\begin{eqnarray}\label{C2}
&&{\mathcal{Q}}A^{a}=-(D^{\mathsf{adj}}c)^{a}+igh^{abd}[A^{d},c^{b}]_{\star}\ ,\\
&&{\mathcal{Q}}\psi=-ig(c^{a}t^{a}\star\psi)\ ,\\
&&{\mathcal{Q}}\bar{c}^{a}=\xi^{-1}(\partial\cdot A^{a})\ ,\\
&&{\mathcal{Q}}c^{a}=\frac{1}{2}\ gf^{abd}c^{b}\star
c^{d}-\frac{i}{4}\ gd^{abd}\{c^{b},c^{d}\}_{\star}\ .
\end{eqnarray}
Now we want to find the action of ${\mathcal{Q}}^{2}$ on each
field of the theory. The reason is that one can easily show:
\begin{eqnarray}\label{CC}
{\mathcal{Q}}^{2}\phi=[Q^{2},\phi]\ ,
\end{eqnarray}
which indicates the relation between ${\mathcal{Q}}^{2}$ and
$Q^{2}$. First, consider the antighost field:
\begin{eqnarray}\label{C3}
{\mathcal{Q}}^{2}\bar{c}^{a}&=&{\mathcal{Q}}({\mathcal{Q}}\bar{c}^{a})\nonumber\\
&=&\xi^{-1}\partial\cdot({\mathcal{Q}}A^{a})\nonumber\\
&=&-\xi^{-1}\partial\cdot\Big((D^{\mathsf{adj}}c)^{a}-igh^{abd}[A^{d},c^{b}]_{\star}\Big)\
.
\end{eqnarray}
Using the equation of motion governed on the antighost field for
the above expression, we immediately conclude:
\begin{eqnarray}\label{C4}
{\mathcal{Q}}^{2}\bar{c}^{a}=0\ .
\end{eqnarray}
Considering Eq. (4.4) and Eq. (4.6), the action of
${\mathcal{Q}}^{2}$ on the matter field yields:
\begin{eqnarray}\label{C5}
{\mathcal{Q}}^{2}\psi&=&{\mathcal{Q}}({\mathcal{Q}}\psi)\nonumber\\
&=&-ig({\mathcal{Q}}c^{a})t^{a}\star\psi+igc^{a}t^{a}\star({\mathcal{Q}}\psi)\nonumber\\
&=&-\frac{1}{2}g^{2}f^{abd}c^{b}\star
c^{d}t^{a}\star\psi-\frac{1}{4}g^{2}d^{abd}\{c^{b},c^{d}\}_{\star}\star
t^{a}\psi+g^{2}c^{a}\star c^{b}(t^{a}t^{b})\star\psi\ ,
\end{eqnarray}
where $c^{a}\star c^{b}(t^{a}t^{b})$ can be written in the form:
\begin{eqnarray}\label{C6}
c^{a}\star c^{b}(t^{a}t^{b})&=&h^{dab}c^{a}\star
c^{b}t^{d}\nonumber\\
&=&\Big(\frac{i}{2}f^{abd}c^{b}\star
c^{d}+\frac{1}{4}d^{abd}\{c^{b},c^{d}\}_{\star}\Big)t^{a}\ ,
\end{eqnarray}
which leads to the vanishing result for ${\mathcal{Q}}^{2}\psi=0$.
For the gauge field, however; we have:
\begin{eqnarray}\label{C8}
{\mathcal{Q}}^{2}A^{a}&=&{\mathcal{Q}}({\mathcal{Q}}A^{a})\nonumber\\
&=&-{\mathcal{Q}}\Big((D^{\mathsf{adj}}c)^{a}-igh^{acb}[A^{b},c^{c}]_{\star}\Big)
\end{eqnarray}
substituting the following equation:
\begin{eqnarray}\label{C9}
{\mathcal{Q}}[A^{b},c^{c}]_{\star}=\{{\mathcal{Q}}A^{b},c^{c}\}_{\star}+[A^{b},{\mathcal{Q}}c^{c}]_{\star}\
,
\end{eqnarray}
for Eq. (\ref{C8}), we arrive at:
\begin{eqnarray}\label{C10}
{\mathcal{Q}}^{2}A^{a}&=&-\Biggr[\partial({\mathcal{Q}}c^{a})+\frac{1}{2}gf^{abc}[{\mathcal{Q}}A^{b},
c^{c}]_{\star}+\frac{1}{2}gf^{abc}\{A^{b},{\mathcal{Q}}c^{c}\}_{\star}\nonumber\\
&&\hspace{2cm}-\frac{i}{2}gd^{abc}\{{\mathcal{Q}}A^{b},c^{c}\}_{\star}-\frac{i}{2}gd^{abc}[A^{b},
{\mathcal{Q}}c^{c}]_{\star}\Biggr]\ .
\end{eqnarray}
Applying Eq. (4.3) and Eq. (4.6) for the above relation, we have:
\begin{eqnarray}\label{C11}
{\mathcal{Q}}^{2}A^{a}&=&-\partial\Big(\frac{1}{2}gf^{abc}c^{b}\star
c^{c}-\frac{i}{4}gd^{abc}\{c^{b},c^{c}\}_{\star}\Big)\nonumber\\
&&+\frac{1}{2}gf^{abc}\Big[(D^{\mathsf{adj}}c)^{b}-igh^{bed}[A^{d},c^{e}]_{\star},c^{c}\Big]_{\star}\nonumber\\
&&-\frac{1}{2}gf^{abc}\Big\{A^{b},\frac{1}{2}gf^{cde}c^{d}\star
c^{e}-\frac{i}{4}gd^{cde}\{c^{d},c^{e}\}_{\star}\Big\}_{\star}\nonumber\\
&&+\frac{i}{2}gd^{abc}\Big[A^{b},\frac{1}{2}gf^{cde}c^{d}\star
c^{e}-\frac{i}{4}gd^{cde}\{c^{d},c^{e}\}_{\star}\Big]_{\star}\nonumber\\
&&-\frac{i}{2}gd^{abc}\Big\{(D^{\mathsf{adj}}c)^{b}-igh^{bed}[A^{d},c^{e}]_{\star},c^{c}\Big\}_{\star}\
.
\end{eqnarray}
Now consider the terms include space-time derivative:
\begin{eqnarray}\label{C12}
\mbox{terms including
derivative}&=&\frac{1}{2}gf^{abc}\partial(c^{b}\star
c^{c})-\frac{i}{4}gd^{abc}\partial\{c^{b},c^{d}\}_{\star}\nonumber\\
&&-\frac{1}{2}gf^{abc}[\partial
c^{b},c^{c}]_{\star}+\frac{i}{2}gd^{abc}\{\partial
c^{b},c^{c}\}_{\star}=0\ .
\end{eqnarray}
For the rest terms, use the following identity:
\begin{eqnarray}\label{C13}
f^{bde}A^{d}\star
c^{e}-ih^{bed}[A^{d},c^{e}]_{\star}=-ih^{bde}A^{d}\star
c^{e}+ih^{bed}c^{e}\star A^{d}\ ,
\end{eqnarray}
we will obtain:
\begin{eqnarray}\label{C14}
{\mathcal{Q}}^{2}A^{a}&=&\frac{1}{2}\
g^{2}(h^{abc}h^{cde}-h^{ace}h^{cbd})A^{b}\star
c^{d}\star c^{e}\nonumber\\
&&\hspace{-0.3cm}+\frac{1}{2}\
g^{2}(h^{adc}h^{ceb}-h^{acb}h^{cde})c^{d}\star c^{e}\star
A^{b}\nonumber\\
&&\hspace{-0.3cm}+\frac{1}{2}\
g^{2}(h^{abe}h^{bcd}-h^{acb}h^{bde})c^{c}\star A^{d}\star c^{e}\ .
\end{eqnarray}
Using the following Jacobi identity (see appendix B.) between
$h^{abc}$'s constants:
\begin{eqnarray}\label{C15}
h^{abe}h^{bcd}-h^{acb}h^{bde}=0\ ,
\end{eqnarray}
we easily find ${\mathcal{Q}}^{2}A^{a}=0$. As the last step,
consider the action of ${\mathcal{Q}}^{2}$ on the ghost field. To
accomplish calculations in an easier way, notice that we can
rewrite Eq. (4.6) as follows:
\begin{eqnarray}\label{C16}
{\mathcal{Q}}c^{a}=-igh^{abd}c^{b}\star c^{d}\ .
\end{eqnarray}
Now using Eq. (4.6), we gain the action of ${\mathcal{Q}}^{2}$ on
the ghost field as:
\begin{eqnarray}\label{C17}
{\mathcal{Q}}^{2}c^{a}&=&{\mathcal{Q}}({\mathcal{Q}}c^{a})\nonumber\\
&=&-igh^{abd}\Big(({\mathcal{Q}}c^{b})\star c^{d}-c^{b}\star({\mathcal{Q}}c^{d})\Big)\nonumber\\
&=&-igh^{abd}\Big(-igh^{bef}c^{e}\star c^{f}\star
c^{d}+igh^{def}c^{b}\star c^{e}\star c^{f}\Big)\nonumber\\
&=&-g^{2}(h^{abd}h^{bef}-h^{aeb}h^{bfd})c^{e}\star c^{f}\star
c^{d}\ ,
\end{eqnarray}
which will vanish if we use the Jacobi identity Eq. (\ref{C15}).
In this manner, we found that:
\begin{eqnarray}\label{C18}
{\mathcal{Q}}^{2}\phi=0\ ,\hspace{1.5cm}
\phi\in\{A^{a},\psi,c^{a},\bar{c}^{a}\}\ .
\end{eqnarray}
Remembering Eq. (\ref{CC}), the above result implies that:
\begin{eqnarray}\label{C19}
[Q^{2},\phi]=0\ .
\end{eqnarray}
For this to be satisfied for all operators $\phi$, it is
necessary for $Q^{2}$ either to vanish or be proportional to the
unit operator. But $Q^{2}$ cannot be proportional to the unit
operator because it has a non-vanishing ghost quantum number, so
it must vanish:
\begin{eqnarray}\label{C20}
Q^{2}=0\ .
\end{eqnarray}
\section{The Space of Physical States and Unitarity}
\setcounter{equation}{0} In keeping with the BRST quantization
method, we found a global symmetry transformation (the BRST
transformation) for noncommutative gauge theory such that its
associated charge is nilpotent. In this section, however, we will
establish the Hilbert space of physical states, following the rest
of our discussion in the Hamiltonian formalism. First, in Sect.
5.1, we discuss that the BRST symmetry leads to a conserved
charge only for the space-like non-commutativity parameter. In
the next part, Sect. 5.2, we make sure that the BRST symmetry
preserves at quantum level. Finally in Sect. 5.3, it turns out
the subspace of physical states is the cohomology space of the
BRST operator.
\subsection{Conserved Charge and Space-like Non-commutativity}
Although the BRST symmetry was proved for a general
non-commutativity parameter, we discuss this symmetry leads to a
conserved charge only for the space-like non-commutativity
parameter ($\theta^{0i}=0$). The rest of our arguments is
therefore restricted to this case.
\par
The common way to obtain the charge of a continuous global
symmetry of the theory is replacing the local function for the
global parameter of the symmetry transformation. In this manner,
there is no necessity for the new transformation to be the
symmetry of the theory, too. In fact, the variation of the
action, under this new transformation, introduces the current
associated to the global symmetry of the theory. Although this
variation vanishes for all local parameters of transformation, we
cannot conclude that the divergence of the current has to vanish,
in contrast to the commutative case. Since the Moyal star product
can be removed in any quadratic term of the action, the most
general result which can be deduced is that the divergence of the
current is equal to the Moyal bracket of some functions\cite{h5}.
The mentioned Moyal bracket disappears only for the space-like
non-commutativity when we integrate on the continuity equation
over all spatial coordinates in order to obtain the time
variation of the charge. Consequently, the charge associated to a
symmetry transformation commutes with the Hamiltonian of the
theory in this case. Hence, the rest of our arguments are
restricted to the space-like non-commutativity.
\subsection{BRST Symmetry at Quantum Level}
So far, we have shown that the full action of the noncommutative
gauge theory, Eq. (\ref{A13}), is invariant under the BRST
transformation, Eq. (\ref{B18}). But in this stage, we cannot
still claim that this symmetry is preserved at quantum level. On
other words, we must show that the path integral measure of the
partition function of the theory remains invariant under the BRST
transformation. The sourceless partition function of the
noncommutative gauge theory is given by:
\begin{eqnarray}\label{D1}
{\mathcal{Z}}=\int {\mathcal{D}}A\ {\mathcal{D}}\psi\
{\mathcal{D}}\bar{\psi}\ {\mathcal{D}}c\ {\mathcal{D}}\bar{c}\
e^{iS[A,\psi,\bar{\psi},c,\bar{c}]}\ .
\end{eqnarray}
One can immediately convince himself that the above path integral
measure is invariant under the BRST transformation. For instance,
the Jacobian describes the change in ${\mathcal{D}}c$ is the
determinant of:
\begin{eqnarray}\label{D2}
\frac{\delta}{\delta c^{b}(y)}\big(c^{a}(x)+\delta
c^{a}(x)\big)=\Big(\delta^{ab}\delta_{y}-ig(\delta\lambda)h^{abd}\delta_{y}\star
c^{d}-ig(\delta\lambda)h^{adb}c^{d}\star\delta_{y}\Big)(x)\ ,
\end{eqnarray}
where we defined $\delta_{y}(x)\equiv\delta(x-y)$. Obviously,
this determinant is equal to one since $(\delta\lambda)^{2}=0$.
Therefore, the partition function of the noncommutative gauge
theory is invariant under the BRST transformation.
\subsection{Hilbert Space of Physical States}
We are already able to establish the subspace of physical states.
Since ghosts violate spin-statistics (being scalar fermions), the
space of states of the theory included ghosts cannot be an actual
space. Further, the state space has a pseudo-Hermitian rather
Hermitian inner product. But it is possible to construct a certain
subspace ${\mathcal{H}}$ which does not include ghosts and has a
Hermitian norm, analogous to the Hilbert space in actual physical
theories.
\par
In order to obtain BRST invariant S-matrix elements, it is
necessary for physical states to belong in the kernel of $Q$ (i.e.
$Q-$closed):
\begin{eqnarray}\label{D3}
Q|\psi\rangle=0\ ,
\end{eqnarray}
just as the commutative case. Since charge $Q$ is nilpotent, each
two physical states that differ only by a state in the image of
$Q$ (i.e. $Q-$exact), have the same matrix elements with all
other physical states, and are therefore physically equivalent.
Hence, we define an equivalence relation to put equivalent states
in the same class. Two states $|\psi\rangle$ and $|\psi'\rangle$
are equivalent: $|\psi'\rangle\sim|\psi\rangle$ if there is some
state $|\chi\rangle$ such that:
\begin{eqnarray}\label{D4}
|\psi'\rangle=|\psi\rangle+Q|\chi\rangle\ .
\end{eqnarray}
The set of equivalent classes is nothing but the cohomology of
$Q$. Thus, the space of physical states is isomorphic to the
quotient space:
\begin{eqnarray}\label{D5}
{\mathcal{H}}\sim\mbox{Ker}Q\Big/\mbox{Im}Q\ .
\end{eqnarray}
\par
As the final discussion, we will prove the unitarity of the
S-matrix elements in the subspace of physical states. Consider a
typical S-matrix element:
\begin{eqnarray}\label{D6}
_{\mbox{\tiny{out}}}\langle\alpha|\beta\rangle_{\mbox{\tiny{in}}}=
\langle\alpha|S^{\dag}S|\beta\rangle\ ,
\end{eqnarray}
where $_{\mbox{\tiny{out}}}\langle\alpha|$ and
$|\beta\rangle_{\mbox{\tiny{in}}}$ are asymptotic states, whereas
$|\alpha\rangle,|\beta\rangle$ are two physical states of the
theory. Since $Q$ commutes with the Hamiltonian of the theory, the
time evolution of any such states must also be annihilated by~$Q$:
\begin{eqnarray}\label{D7}
Q\ S|\beta\rangle=0\ .
\end{eqnarray}
The above expression states that $S|\beta\rangle$ is a linear
combination of states in $\mbox{Ker}Q$. On other words, we have:
\begin{eqnarray}\label{D78}
S|\beta\rangle=\sum_{|\gamma\rangle\in{\mathcal{P}}_{1}}|\gamma\rangle\langle\gamma|S|\beta\rangle\
,\hspace{2cm}{\mathcal{P}}_{1}=\big\{|\gamma\rangle:\
\mbox{\small{Ker}}Q=\mbox{\small{span}}\{|\gamma\rangle\}\big\}\ ,
\end{eqnarray}
where ${\mathcal{P}}_{1}$ is a basis for subspace $\mbox{Ker}Q$.
Again for the states, in the kernel of $Q$, can be written in the
form $|\gamma\rangle+Q|\chi\rangle$, we find:
\begin{eqnarray}\label{D9}
\Big(\langle\gamma|+\langle\chi|Q\Big)S|\beta\rangle=\langle\gamma|S|\beta\rangle+
\langle\chi|QS|\beta\rangle\ ,
\end{eqnarray}
but the second term of the r.h.s. of the above equality vanishes
since $[Q,H]=0$. This result implies that:
\begin{eqnarray}\label{D10}
S|\beta\rangle=\sum_{|\gamma\rangle\in{\mathcal{P}}_{2}}|\gamma\rangle\langle\gamma|S|\beta\rangle\
,\hspace{2cm}{\mathcal{P}}_{2}=\big\{|\gamma\rangle:\
{\mathcal{H}}=\mbox{\small{span}}\{|\gamma\rangle\}\big\}\ ,
\end{eqnarray}
where ${\mathcal{P}}_{2}$ is a basis for the Hilbert space of
physical states. This relation guarantees the unitarity of
S-matrix elements for the subspace of physical states.
\section{Conclusion}
In this paper, the BRST quantization method is followed for
noncommutative gauge theories. Of course, our discussions are
presented in the framework of the Moyal noncommutative gauge
theory. Nevertheless, all obtained results are valid for any
noncommutative gauge theory whose $C^{*}$ algebra of functions on
${\mathcal{R}}^{d}$ is associative with respect to its star
product and also the trace defined on this algebra satisfies the
cyclicity property.
\par
After introducing the full action of the Moyal noncommutative
$U(N)$ gauge theory in Sect. 2, however, the BRST symmetry
transformation is presented for this theory in Sect. 3. In
analogy to the commutative case, the BRST transformations for the
gauge and matter fields are nothing but infinitesimal gauge
transformations. Moreover, as expected, this transformation will
take its commutative form if we put $\theta=0$.
\par
In Sect. 4, the nilpotency of the charge associated to the BRST
symmetry is then proved. Each nilpotent operator which commutes
with the Hamiltonian of the theory has many useful advantages.
But in the first part of Sect. 5, it is proved the charge
associated to a continuous global symmetry of the theory is not
conserved in general, in contrast to the commutative case. Due to
the cyclicity property of the trace with respect to the star
product, the associated charge is conserved only for the
space-like non-commutativity parameter. In this case, the expected
commutation relation between the charge and the Hamiltonian of
the theory is restored.
\par
In the next part of this section, Sect. 5.2, the BRST symmetry
transformation is considered at quantum level. To preserve this
transformation as a symmetry at quantum level, it is necessary
for the the path integral measure of the partition function to
remain invariant under the BRST transformation. This is easily
shown since the global parameter of the BRST transformation is a
Grassmann variable.
\par
Since the ghost and antighost fields violate the spin-statistics
theorem, the space of states included ghosts is not a Hilbert
space. Hence, in the last part of Sect. 5, the BRST cohomology is
taken to produce the physical Hilbert space just as the
commutative case. Ultimately, it is proved that the S-matrix
elements projected onto the physical space of states are unitary.
\par
Since the BRST symmetry is a continuous transformation, it
generates a set of Ward identities for noncommutative non-Abelian
gauge theories. Therefore, the BRST symmetry will provide a
powerful tool to study the renormalization properties of
noncommutative gauge theories in \textit{all orders} of the
perturbative expansion.
\section{Acknowledgement}
The author thanks the theory group of physics department of
Sharif University of Technology.
\newpage
\begin{appendix}
\setcounter{equation}{0}
\section{More Details to Obtain Eq. (\ref{B16})}
In this section, we present more details for calculations that
lead to Eq. (\ref{B16}). First, let us extract term
$\frac{i}{4}g^{2}f^{abc}d^{cde}\{c^{d},c^{e}\}_{\star}\delta\lambda$
which is appeared in the second line of Eq. (\ref{B16}). To do
this, consider the last term in the r.h.s. of Eq. (\ref{B13})
which is:
\begin{eqnarray}\label{E1}
\frac{i}{2}g^{2}f^{abc}d^{bed}[A^{d},c^{e}]_{\star}\star
c^{c}\delta\lambda=\frac{i}{2}g^{2}f^{abc}d^{bed}(A^{d}\star
c^{e}\star c^{c}-c^{e}\star A^{d}\star c^{c})\delta\lambda\ .
\end{eqnarray}
We use the following Jacobi identity (see appendix B.) for the
r.h.s. of the above relation:
\begin{eqnarray}\label{E2}
f^{abc}d^{bed}=f^{aeb}d^{bcd}-f^{abd}d^{ceb}\ ,
\end{eqnarray}
the fist term in the r.h.s. of Eq. (\ref{E1}) itself becomes two
terms that one of them is just we need. Up to now, the remained
terms only for the first line of Eq. (\ref{B12}) are:
\begin{eqnarray}\label{E3}
&&\frac{1}{2}g^{2}f^{abc}f^{bed}c^{e}\star A^{d}\star
c^{c}\delta\lambda-\frac{1}{2}g^{2}f^{abe}f^{bdc}A^{d}\star
c^{e}\star c^{c}\delta\lambda\nonumber\\
&&\hspace{-0.3cm}-\frac{i}{2}g^{2}f^{abc}d^{bed}c^{e}\star
A^{d}\star
c^{c}\delta\lambda+\frac{i}{2}g^{2}f^{aeb}d^{bcd}A^{d}\star
c^{e}\star c^{c}\delta\lambda\ ,
\end{eqnarray}
which can be written in the form:
\begin{eqnarray}\label{E4}
-ig^{2}f^{abc}h^{bed}\Big\{A^{d}\star
c^{c},c^{e}\Big\}_{\star}\delta\lambda\ .
\end{eqnarray}
We release this remained term from the first line of Eq.
(\ref{B12}) for a moment, and pay attention to the second line of
Eq. (\ref{B12}). We want to form terms are appeared in the last
line of Eq. (\ref{B16}). Considering the second term of the
second line of Eq. (\ref{B12}):
\begin{eqnarray}\label{E5}
g^{2}h^{abd}h^{dfe}\Big\{[A^{e},c^{f}]_{\star},c^{b}\Big\}_{\star}\delta\lambda=
g^{2}h^{abd}h^{dfe}\Big(\Big\{A^{e}\star
c^{f},c^{b}\Big\}_{\star}-\Big\{c^{f}\star
A^{e},c^{b}\Big\}_{\star}\Big)\delta\lambda\ ,
\end{eqnarray}
where we rewrite its first term in the r.h.s. as follows:
\begin{eqnarray}\label{E6}
g^{2}h^{abd}h^{dfe}\Big\{A^{e}\star
c^{f},c^{b}\Big\}_{\star}=g^{2}h^{abd}(\frac{i}{2}f^{dfe}+\frac{1}{2}d^{dfe})\Big\{A^{e}\star
c^{f},c^{b}\Big\}_{\star}\ .
\end{eqnarray}
Adding the first term of the r.h.s. of the above equality to the
first term of the second line of Eq. (\ref{B12}), we obtain:
\begin{eqnarray}\label{E7}
\frac{i}{2}g^{2}h^{abd}f^{def}\Big\{A^{e}\star
c^{f},c^{b}\Big\}_{\star}\delta\lambda\ .
\end{eqnarray}
Rewrite the second term of Eq. (\ref{E5}) as Eq. (\ref{E6}) and
then add its $h^{abd}f^{dfe}$ term to the above result (Eq.
(\ref{E7})), we gain:
\begin{eqnarray}\label{E8}
\frac{i}{2}g^{2}h^{abd}f^{def}\Big\{\{A^{e},c^{f}\}_{\star},c^{b}
\Big\}_{\star}\delta\lambda\ .
\end{eqnarray}
Using the following Jacobi identity between $h^{abc}$ and
$f^{abc}$ constants (see appendix B.):
\begin{eqnarray}\label{E9}
h^{abd}f^{dce}+h^{dbe}f^{cad}-h^{dea}f^{bcd}=0\ ,
\end{eqnarray}
for Eq. (\ref{E8}), we obtain:
\begin{eqnarray}\label{E10}
\frac{i}{2}g^{2}h^{abd}f^{def}\Big\{\{A^{e},c^{f}\}_{\star},c^{b}
\Big\}_{\star}\delta\lambda=\frac{i}{2}g^{2}(-h^{ade}f^{dbf}+f^{adf}h^{dbe})\Big\{\{A^{e},c^{f}\}_{\star},c^{b}
\Big\}_{\star}\delta\lambda\ ,
\end{eqnarray}
where its first term,
$-igh^{ade}\Big[A^{e},-\frac{1}{2}gf^{dfb}c^{f}\star
c^{b}\delta\lambda\Big]_{\star}$, is exactly one of the terms has
been appeared in the third line of Eq. (\ref{B16}). To find the
last term in the third line of Eq. (\ref{B16}), we add the second
term in the r.h.s. of Eq. (\ref{E6}) to the similar term
($f^{abd}d^{dfe}$ term) derived from the second term of the r.h.s.
of Eq. (\ref{E5}) :
\begin{eqnarray}\label{E12}
\frac{1}{2}g^{2}h^{abd}d^{dfe}\Big(\Big\{A^{e}\star c^{f},c^{b}
\Big\}_{\star}-\Big\{c^{f}\star
A^{e},c^{b}\Big\}_{\star}\Big)\delta\lambda\ .
\end{eqnarray}
and use the following Jacobi identity (see appendix B.):
\begin{eqnarray}\label{E13}
h^{abd}d^{dfe}-h^{ade}d^{dbf}-if^{fad}h^{dbe}=0\ ,
\end{eqnarray}
we immediately find Eq. (\ref{E12}) as follows:
\begin{eqnarray}\label{E14}
\frac{1}{2}g^{2}(h^{ade}d^{dbf}-if^{afd}h^{dbe})\Big\{[A^{e},c^{f}]_{\star}
,c^{b}\Big\}_{\star}\delta\lambda\ .
\end{eqnarray}
The first term of the above expression:
\begin{eqnarray}\label{E15}
-igh^{ade}\Big[A^{e},\frac{i}{4}gd^{dbf}\{c^{b},c^{f}\}_{\star}\Big]_{\star}\
,
\end{eqnarray}
is just the needed term. As  the final task, we must show that
the all rest terms cancel themselves. All remained terms from the
second line of Eq. (\ref{B12}) are the second term in the r.h.s.
of Eq. (\ref{E10}) and the second term of Eq. (\ref{E14}). The sum
of these two terms is:
\begin{eqnarray}\label{E16}
ig^{2}f^{adf}h^{dbe}\Big\{A^{e}\star
c^{f},c^{b}\Big\}_{\star}\delta\lambda=ig^{2}f^{abc}h^{bed}\Big\{A^{d}\star
c^{c},c^{e}\Big\}_{\star}\delta\lambda\ .
\end{eqnarray}
This result is just minus of what we obtained in Eq. (\ref{E4}).
Therefore, the rest term of the first line of Eq. (\ref{B12})
cancels the rest term of the second line. Hence, the BRST
transformation which introduced by Eq. (\ref{B18}) indeed remains
the action of the theory (Eq. (\ref{A10})) invariant and
therefore is the symmetry of the theory.
\newpage
\section{Identities Between the Structure Constants of $U(N)$}
\setcounter{equation}{0} \vspace{1cm}
\begin{itemize}
\item Jacobi identities between $f^{abc}$ and $d^{abc}$ are:
\begin{eqnarray*}
&f^{abd}f^{bce}+f^{cba}f^{bde}+f^{dbc}f^{bae}=0\ ,\hspace{1.5cm}
d^{abd}d^{bce}-d^{cba}d^{bde}+f^{dbc}f^{bae}=0\ ,\nonumber\\
&d^{abd}d^{bce}-f^{cba}f^{bde}-d^{dbc}d^{bae}=0\ ,\hspace{1.5cm}
f^{abd}f^{bce}+d^{cba}d^{bde}-d^{dbc}d^{bae}=0\ ,\nonumber\\
&f^{abd}d^{bce}+f^{cba}d^{bde}+f^{dbc}d^{bae}=0\ ,\hspace{1.5cm}
f^{abd}d^{bce}-f^{cba}d^{bde}-d^{dbc}f^{bae}=0\ ,\nonumber\\
&f^{abd}d^{bce}+d^{cba}f^{bde}-f^{dbc}d^{bae}=0\ ,\hspace{1.5cm}
d^{abd}f^{bce}-f^{cba}d^{bde}+f^{dbc}d^{bae}=0\ .\nonumber
\end{eqnarray*}
\item Jacobi identities between $h^{abc}$ and $f^{abc}$ can easily be
obtained by combining above relations:
\begin{eqnarray*}
&h^{abd}f^{bce}+h^{cba}f^{bde}+h^{dbc}f^{bae}=0\ ,\hspace{1.5cm}
f^{abd}h^{bce}+h^{cba}f^{bde}-f^{dbc}h^{bea}=0\ ,\nonumber\\
&f^{abd}h^{bce}-f^{cba}h^{bed}-h^{dcb}f^{bae}=0\ ,\hspace{1.5cm}
h^{abd}f^{bce}-f^{cba}h^{bed}+f^{dbc}h^{bae}=0\ .\nonumber
\end{eqnarray*}
\item Jacobi identities between $h^{abc}$ and $d^{abc}$ can also be obtained in a similar way:
\begin{eqnarray*}
&h^{abd}d^{bce}+ih^{cba}f^{bde}-h^{dbc}d^{bae}=0\ ,\hspace{1.5cm}
if^{abd}h^{bec}+d^{cba}h^{bde}-h^{dbc}d^{bae}=0\ ,\nonumber\\
&h^{abd}d^{bce}-h^{cab}d^{bde}+if^{dbc}h^{bea}=0\ ,\hspace{1.5cm}
h^{abd}d^{bce}+if^{cba}h^{bde}-h^{dcb}d^{bae}=0\ .\nonumber
\end{eqnarray*}
\item Applying the above various Jacobi identities, we can prove
an identity only for $h^{abc}$'s:
\begin{eqnarray*}
h^{abe}h^{bcd}-h^{acb}h^{bde}=0\ .
\end{eqnarray*}
\end{itemize}
\end{appendix}

\end{document}